\documentclass[letterpaper, 10 pt, conference]{ieeeconf}  

\IEEEoverridecommandlockouts                              

\usepackage{mathptmx} 
\usepackage{amsmath} 
\usepackage{amssymb}  
\usepackage{breqn}
\usepackage{graphicx} 
\usepackage{amsmath,amssymb}
\usepackage{lipsum}
\usepackage{mathtools}
\usepackage{cuted}
\DeclareMathOperator{\E}{\mathbb{E}}%

\title{\LARGE \bf
Monocular Retinal Depth Estimation and Joint Optic Disc and Cup Segmentation using Adversarial Networks 
}

\author{Sharath M Shankaranarayana$^{1}$ and Keerthi Ram$^{2}$ and Kaushik Mitra$^{1}$ and Mohanasankar Sivaprakasam$^{1,2}$
\thanks{$^{1}$Dept of Electrical Engineering, IIT-Madras}%
\thanks{$^{2}$Healthcare Technology Innovation Centre, IIT-Madras}%
}

\begin{document}

\maketitle
\thispagestyle{empty}
\pagestyle{empty}

\begin{abstract}
One of the important parameters for the assessment of glaucoma is optic nerve head (ONH) evaluation, which usually involves depth estimation and subsequent optic disc and cup boundary extraction. Depth is usually obtained explicitly from imaging modalities like optical coherence tomography (OCT) and is very challenging to estimate depth from a single RGB image. To this end, we propose a novel method using adversarial network to predict depth map from a single image. The proposed depth estimation technique is trained and evaluated using individual retinal images from INSPIRE-stereo dataset. We obtain a very high average correlation coefficient of 0.92 upon five fold cross validation outperforming the state of the art. We then use the depth estimation process as a proxy task for joint optic disc and cup segmentation.  
\end{abstract}

\section{INTRODUCTION}
\label{sec:intro}
Glaucoma is one of the serious vision threatening ocular disorders, where there is a gradual degeneration in the optic nerve head (ONH) of the retina. Screening is done using 2D fundus imaging for the assessment of optic disc (OD) and cup. Vertical cup to disc ratio (CDR), which is a quantitative measure for measuring the enlargement of cup with respect to disc, is an important indicator of the disease and requires accurate delineation of OD and the cup, which is typically done by a skilled grader. 

There have been many works on optic disc and cup segmentation. and \cite{survey} provides a survey of different techniques. Many methods based on morphological techniques \cite{morphology} and deformable energy based models \cite{related2}\cite{related3} and graph cuts \cite{related6} have been proposed. Recently, upon the advent of deep learning, the U-net \cite{unet} like fully convolutional architectures have been used for many kinds of semantic segmentation task. For the case of glaucoma, recently \cite{cmig} proposed the use CNNs where filters are learned in a greedy fashion and the image is passed through the CNN to get pixelwise predictions for disc and cup segmentation. We recently proposed an end to end fully convolutional network for the task of joint optic disc and cup segmentation \cite{MyPaper}. The work also explored the use of adversarial training for segmentation task.

Depth is also an important cue for assessment of glaucoma and explicitly measuring depth requires complicated imaging techniques such as stereoscopic imaging or optical coherence tomography (OCT). But using these modalities at large scale is infeasible due to their cost, difficulty in operating and portability. This necessitates the need of a method for depth estimation from a single image. Single image depth estimation is highly challenging task and has been explored using deep learning \cite{NIPsdepth} \cite{neuralfields} and the authors of \cite{DepthSemSeg} perform not only depth estimation, but also surface normal estimation and semantic segmentation with a common architecture. In the case of retinal imaging, there have been a few works for depth estimation. A method for estimating depth from stereo is proposed in \cite{Multi-Stereo}. Single image depth estimation using a coupled sparse dictionary based supervision method is proposed in \cite{DepthDictionary}. A fast marching based depth estimation is proposed in \cite{DepthAkshaya}. Most of these single image retinal depth estimation methods rely predominantly on the image intensities and hence not fairly robust, thus necessitating the need of a robust method.  

The main contribution of this work are follows
\begin{enumerate}
\item We propose a new scheme for depth estimation of monocular fundus images using a fully convolutional network.
\item We also explore the effectiveness of depth estimation as a proxy task for joint optic disc and cup segmentation. Since the cupping phenomenon occurs in the optic nerve head (ONH), leading to the variations in depth in ONH, the task of depth estimation could naturally serve as a pretraining method for segmentation.
\end{enumerate}
To the best of our knowledge this is the first work addressing monocular retinal depth estimation using a deep learning.
\begin{figure}[tp]
\begin{minipage}{1.0\linewidth}
  \centering
  \centerline{\includegraphics[width=8.5cm]{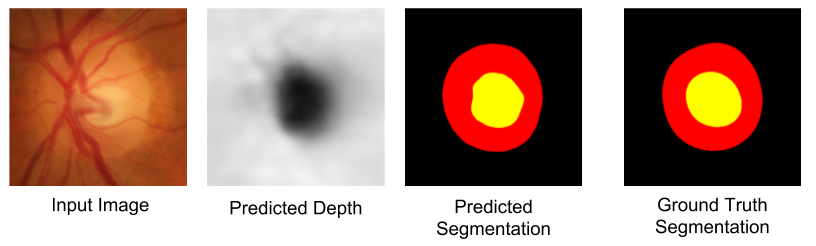}}
  \vspace{-0.350cm}
    \caption{Sample Results from our method}
\vspace{-0.250cm}    
\end{minipage}
\vspace{-0.40cm}
\end{figure}
\section{Methods}
\label{sec:Methods}

Our work consists of two main parts- depth estimation and joint optic disc-cup segmentation. For both these tasks, we employ a fully convolutional network architecture proposed by us in \cite{MyPaper}. The proposed network in \cite{MyPaper} is a generative adversarial network (GAN) based architecture consisting of a generator and a discriminator. The generator is a U-net \cite{unet} like encoder-decoder type architecture with residual connections. The discriminator has a standard CNN architecture employed for classification. The readers are advised to refer \cite{MyPaper} for more details about the architecture. In our set-up, the generator is tasked with learning the required mappings for both the cases of depth estimation and joint optic disc-cup segmentation.
\begin{figure*}[tp]
\centering
\includegraphics[width=\textwidth]{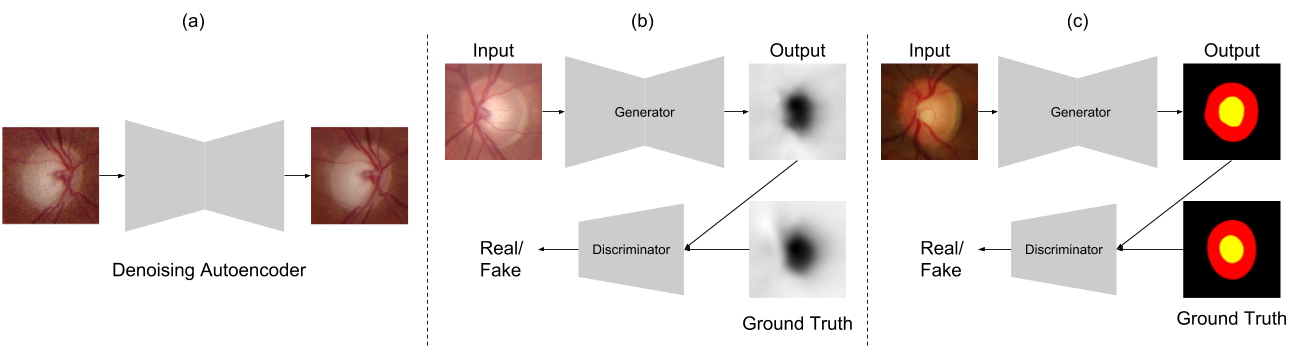}
  \vspace{-0.50cm}
\caption{Our proposed framework: the first part (a) consists of pretraining using a denoising autoencoder which serves as weight initialization for part (b), which consists of depth estimation and also serves as weight initialization for part (c), which consists of segmentation}
\end{figure*}

Single image depth estimation is a challenging task and more so in our case because of the unavailability of a large dataset. Hence, we first collect retinal images of various datasets and crop the region of interest, which in this case is the region around optic disc. With this dataset and with the same architecture as the generator,  we train a denoising autoencoder in order to learn the retinal representations. Thus, we pretrain the generator part of the network as a denoising autoencoder. Once pretrained,  we proceed with depth estimation and disc-cup segmentation (refer figure Fig. 2) as described in the subsequent subsections. 
\subsection{Depth Estimation}
For the first task, our goal is to train a fully convolutional network to predict depth from a single RGB fundus image. For this, we employ the pretrained generator network to learn a mapping between the fundus image and the corresponding depth map. We solve for depth estimation as a  regression problem. Given an RGB fundus image $I$ and the corresponding depth map $d$, our network learns the mapping $G_{depth}(I)$. As in the case with any regression problem, we employ the standard $L_2$ loss function. 
\begin{dmath}\label{eq1}
L_{L2}(G_{depth})=\E_{I,d\sim p_{data}(I,d)}[\|(d - G_{depth}(I)\|_2]
\end{dmath}
Additionally, we also augment $L_2$ loss with adversarial loss so as to improve the depth estimation. 
\begin{dmath}\label{eq2}
G_{depth}^* = arg \underset{G_{depth}}{\mathrm{min}} \underset{D_{depth}}{\mathrm{max}}( L_{GAN}(G_{depth},D_{depth}) + \lambda(L_{L2}(G_{depth}))
\end{dmath}
where $D_{depth}$ is the discriminator network for depth and $L_{GAN}$ is the adversarial loss given by:
\begin{dmath}\label{eq3}
L_{GAN}(G_{depth},D_{depth})=\E_{d\sim p_{data}(d)}[log(D_{depth}(d))] +
               \E_{I\sim p_{data}(I)}
               [log (1-D_{depth}(G_{depth}(I))]
\end{dmath}

\subsection{Joint Optic Disc and Cup Segmentation}
The phenomena of cupping leads to an increase in the relative depth between the cup and the disc. This serves as the main motivation for us to explore depth estimation as a proxy task for joint optic disc-cup segmentation. Our goal in this task is to train a fully convolution network for the task of segmentation. We use a GAN based framework where the generator $G_{segment}$ is tasked with learning a mapping between an RGB image as input $x$ and the corresponding segmentation map $y$. The generator needs to produce outputs so as to fool an adversarially trained discriminator $D_{segment}$ which is trained to discriminate between generated segmentation map and real segmentation map. The final objective function for segmentation is given by -  
\begin{dmath}\label{eq4}
G_{segment}^* = arg \underset{G_{segment}}{\mathrm{min}} \underset{D_{segment}}{\mathrm{max}}( L_{GAN}(G_{segment},D_{segment}) + \lambda(L_{L1}(G_{segment}))
\end{dmath}
 where $L_{GAN}$ and $L_{L1}$ are adversarial loss and $L_1$ loss functions given by - 
\begin{dmath}\label{eq5}
L_{GAN}(G_{segment},D_{segment})=\E_{y\sim p_{data}(y)}[log(D_{segment}(y))] +
               \E_{x\sim p_{data}(x)}
               [log (1-D_{segment}(G_{segment}(x))]
\end{dmath}
\begin{dmath}\label{eq6}
L_{L1}(G_{segment})=\E_{x,y\sim p_{data}(x,y)}[\|(y - G_{segment}(x)\|_1]
\end{dmath}
It is to be noted that in the equations \eqref{eq1} and \eqref{eq3} and also \eqref{eq5} and \eqref{eq6}, $p_{data}(x)$ corresponds to the distribution of $x$ and the expectation of the log-likelihood of the pair $(x,y)$ being sampled from the underlying probability distribution of real pairs $p_{data}(x,y)$ is represented by $\E_{x,y\sim p_{data}(x,y)}$.

Since we proposed depth estimation as a proxy task, we initialize the weights of the generator $G_{segment}$ with the weights of fully trained network $G_{depth}$ which was trained for depth estimation.

\section{Experiments and Results}
\begin{figure*}[tp]
\centering
\includegraphics[width=\textwidth]{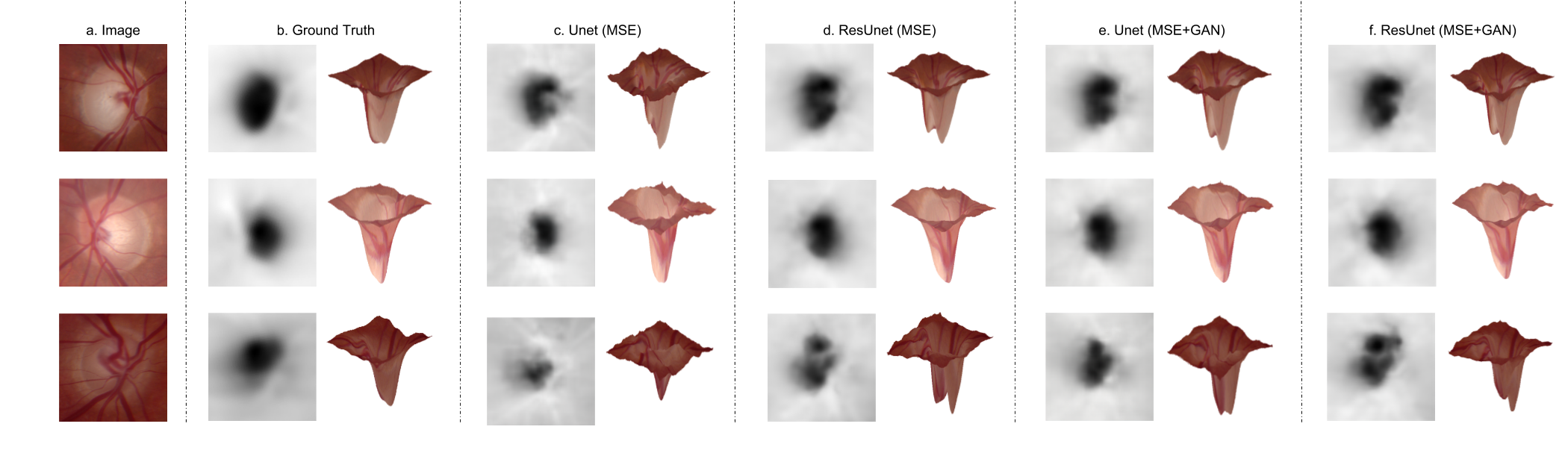}
\vspace{-0.750cm}
\caption{Qualitative results for depth estimation with the input image and depth maps and corresponding surface reconstruction}

\end{figure*}

Since the first task in our work is to perform pretraining using a denoising autoencoder, we first collect retinal images from various sources such as RIMONE, MESSIDOR, DRIVE, STARE etc. We then crop the OD region and add noise to the images and train a denoising autoencoder with different generator architectures such as Unet\cite{unet} and ResUnet\cite{MyPaper}. 

We then use the pretrained deep networks for estimating depth. We use the INSPIRE-stereo dataset \cite{Multi-Stereo} for estimating the depth. The dataset consists of color fundus images along with the ground truth depth map obtained from OCT. 
We use five fold cross-validation which in our case turns out to be $6$ validation images and $24$ training images. We do very heavy data augmentation on the training set with various levels of zoom, gamma jitter along with standard techniques such as flips and rotations, inturn blowing up the training data by a factor of $100$ to aid in training a deep network. We then train the network for $200$ epochs with $Adam$ optimizer. The results of depth estimation are shown in figure 3. The metrics used for quantitative evaluation of depth maps are root mean squared error (RMSE) given by\\
$\sqrt{\Sigma(x_i - y_i)^2}$\\
and  correlation coefficient $r$ given by \\
$r(x,y) = \frac{\Sigma(x_i - \bar{x})(y_i - \bar{y})}{\sqrt{\Sigma(x_i - \bar{x})^2\Sigma(y_i - \bar{y})^2}}$ \\
where $x$ and $y$ are estimated depth maps and ground truth depth maps respectively, and $i$ is the pixel index. We enlist the values obtained for four experiments-\\
U-net with only $L_2$ loss (U-net)\\
Residual U-net with only $L_2$ loss (ResU-net)\\
U-net with adversarial loss (U-GAN)\\
Residual U-net with adversarial loss (ResU-GAN)\\
The values obtained are listed in Table. 1. It can be seen from the table that deep learning based methods yield superior results compared to the other methods for monocular depth estimation. 
\begin{table}
\caption{Comparison of various Depth estimation methods }
  \centering
  \begin{tabular}{|c|c|c|c|c|}
    \hline
    {\textbf{Method}} & \multicolumn{2}{c|}{\textbf{Correlation Coefficient}} & \multicolumn{2}{c|}{\textbf{RMSE}}\\
    \cline{2-5}
    & Mean & Std-Dev & Mean & Std-Dev\\
    \hline
     \cite{Multi-Stereo} & - & - & 0.1592 & 0.0879 \\ \hline
     \cite{DepthDictionary} & 0.8000 & 0.1200 & - & - \\ \hline
     \cite{DepthAkshaya} &  0.8225 & - & 0.1532 & 0.1206 \\ \hline
     Proposed U-net & 0.8322 & 0.1077 & 0.0190 & 0.0094 \\ \hline
     Proposed ResU-net & 0.8984 & 0.0698 & 0.0124 & 0.0079 \\ \hline
     Proposed U-GAN & 0.9268 & \textbf{0.0377} & 0.0105 & 0.0080 \\ \hline
     Proposed ResU-GAN & \textbf{0.9269} & 0.0434 & \textbf{0.0099} & \textbf{0.0054}\\ \hline
 \end{tabular}
  \end{table}
For the task of joint optic disc and cup segmentation, we use the RIM-ONE dataset containing $159$ labeled images for optic disc and cup. Instead of training from scratch, we use the respective depth pretrained networks for  weight initialization. Accordingly, we again have the four experiments for segmentation-\\
Depth pretrained U-net and ResU-net without adversarial loss (DP U-net and DP ResU-net respectively) and depth pretrained U-net and ResU-net with adversarial loss (DP U-GAN and DP ResU-GAN respectively).\\
The delineated outputs can be seen in figure Fig. 4, and the quantitative metrics employed for semantic segmentation are F-score and Intersection over Union (IOU) measures. The values are tabulated in table 2. 

It is interesting to note that depth pretraining leads to improved segmentation accuracy in the case of U-net compared to the network trained from scratch. For the case of ResU-net, it leads to similar performance but pretraining leads to consistent results in the cases of adversarial training and without adversarial training. Perhaps, one of the reasons for depth pretrained models not giving significantly better results compared models trained from scratch could be the dataset bias since the RIMONE dataset and INSPIRE-stereo dataset seem to differ significantly in terms of quality and luminance and color distribution when examined visually. Also, it can be seen from figure Fig. 4 that the depth estimation for RIMONE dataset doesn't seem to yield very accurate results. Also, in-availability of a large dataset for depth estimation could also be one of the causes for not outperforming the models trained from scratch.

\begin{table}
\caption{Comparison of various segmentation methods}
  \centering
  \begin{tabular}{|c|c|c|c|c|}
     \hline
    {\textbf{Method}} & \multicolumn{2}{c|}{\textbf{Optic Disc}} & \multicolumn{2}{c|}{\textbf{Optic Cup}}\\
    \cline{2-5}
    & F-Measure & IOU & F-Measure & IOU\\
    \hline
    \cite{morphology} & 0.901 & 0.842 & - & - \\ \hline
    \cite{related4} & 0.931 & 0.880 & 0.801 & 0.764 \\ \hline
    \cite{newref} & 0.892 & 0.829 & 0.744 & 0.732 \\ \hline
    \cite{cmig} & 0.942 & 0.890 & 0.824 & 0.802  \\ \hline
    U-net \cite{MyPaper} & 0.973 & 0.886 & 0.927 & 0.749 \\ \hline
    U-GAN \cite{MyPaper} & 0.984 & 0.949 & 0.779 & 0.675 \\ \hline
    ResU-net \cite{MyPaper} & 0.977 & 0.901 & 0.945 & 0.786 \\ \hline
    ResU-GAN \cite{MyPaper} & 0.987 & 0.961 & 0.906 & 0.739 \\ \hline
    DP U-net & 0.9841 & 0.9472 & 0.9347 & 0.7395\\ \hline
    DP U-GAN & 0.9841 & 0.9497 & 0.9285 & 0.7390\\ \hline
    DP ResU-net & 0.9857 & 0.9575 & 0.9354 & 0.7458\\ \hline
    DP ResU-GAN & 0.9861 & 0.9575 & 0.9354 & 0.7488 \\ \hline

  \end{tabular}
  \end{table}
\begin{figure*}[tp]
\centering
\includegraphics[width=\textwidth]{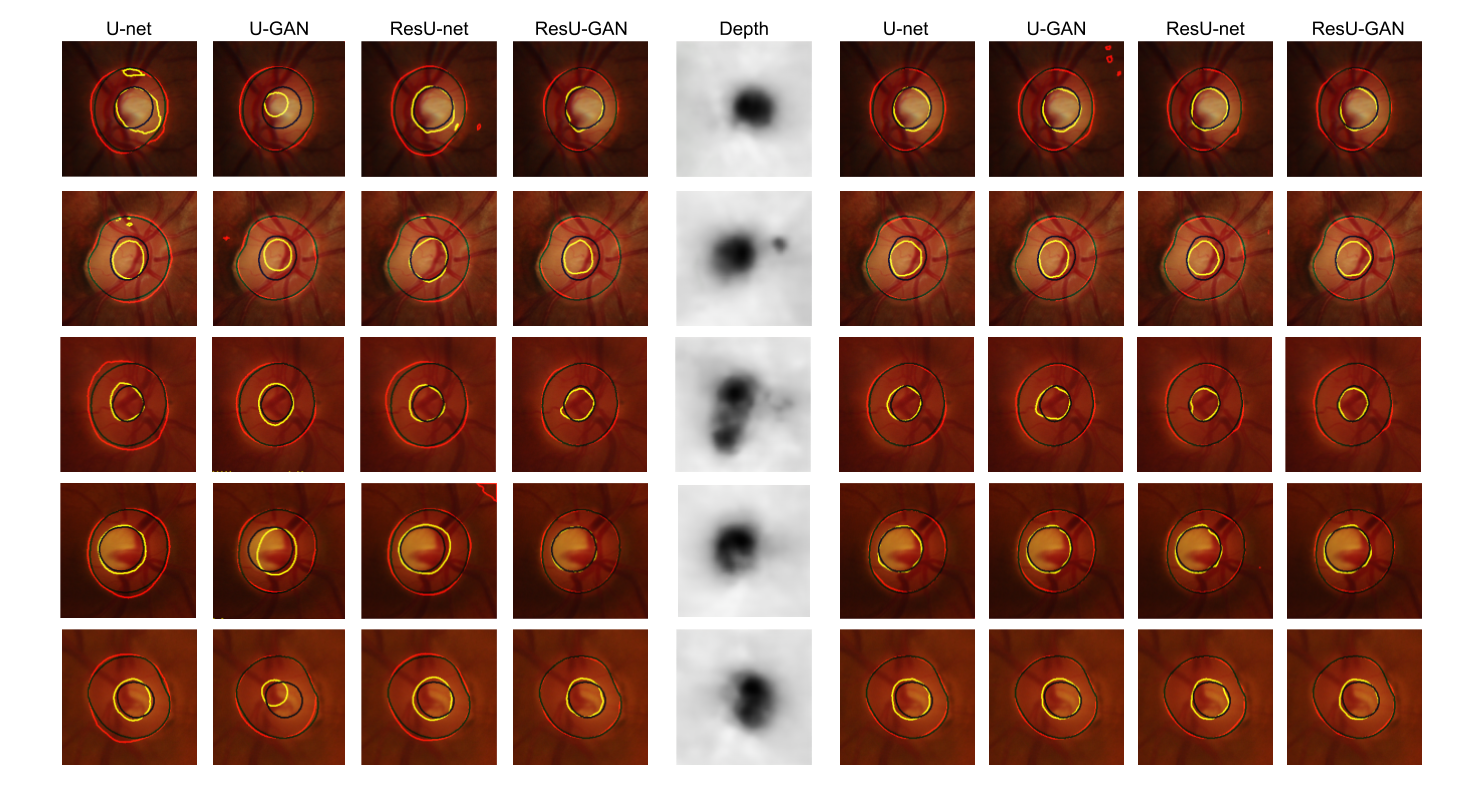}
 \vspace{-0.750cm}
\centering
    \caption{Results for segmentation: the first four columns show the results for network trained from scratch and the last four columns show the results for depth pretrained networks while the middle column displays the estimated depth map for RIMONE dataset}

\end{figure*}

\section{CONCLUSION}
In this work, we proposed a new method for monocular retinal depth estimation using deep learning. It was seen that although this method outperforms other existing methods for the depth estimation by large margin in terms of the usual metrics, its generalization ability is one of the main concerns. The study of depth training as as a proxy task for joint optic disc-cup segmentation highlights the issue of generalization ability. One of the ways to address this would be to use better augmentation techniques to remove the dataset bias. In future, we would also like to explore methods for using the depth information explicitly for semantic segmentation.

\addtolength{\textheight}{-2cm}   




\end{document}